\documentclass[aps,prd,12pt,nofootinbib]{revtex4}

\usepackage{amsfonts}
\usepackage{amsmath,epsfig,bm}
\usepackage{graphicx}
\usepackage{color}

\usepackage{tabularx} 

\usepackage{times}

\newcommand{\be}{\begin{equation}}
\newcommand{\ee}{\end{equation}}
\newcommand{\ba}{\begin{eqnarray}}
\newcommand{\ea}{\end{eqnarray}}
\newcommand{\bd}{\begin{displaymath}}
\newcommand{\ed}{\end{displaymath}}

\def\thalf{{\textstyle{\frac{1}{2}}}}
\def\tthalf{{\textstyle{\frac{3}{2}}}}
\def\oneth{{\textstyle{\frac{1}{3}}}}

\def\oneqt{{\textstyle{\frac{1}{4}}}}

\def\root{\sqrt{-g}}

\def\rt3{\sqrt{3}}
\def\rt6{\sqrt{6}}
\def\mL2{(m_{\phi}L)^2}
\def\Dr{\frac{d}{dr}}

\newcommand{\cS}{\mathcal S}

\begin{document}

\title{{\bf Dynamical AdS/Yang-Mills Model}}
\author{Sean P. Bartz}
\affiliation{Department of Physics and Astronomy, Macalester College, St. Paul, Minnesota 55105,USA}
\author{Aditya Dhumuntarao}
\affiliation{Perimeter Institute for Theoretical Physics, Waterloo, Ontario N2L 2Y5, Canada \\
School of Physics and Astronomy, University of Minnesota, Minneapolis, Minnesota 55455,USA}
\author{Joseph I. Kapusta}
\affiliation{School of Physics and Astronomy, University of Minnesota, Minneapolis, Minnesota 55455,USA}

\vspace{.3cm}
\date{\today}

\parindent=20pt

\begin{abstract}

The AdS/CFT correspondence has provided new and useful insights into the nonperturbative regime of strongly coupled gauge theories. 
We construct a class of models meant to mimic Yang-Mills theory using the superpotential method.  This method allows us to efficiently address the problem of solving all the equations of motion.  The conformal symmetry is broken in the infrared by a dilaton field.  Using a five-dimensional action we calculate the mass spectrum of scalar glueballs.  This spectrum contains a tachyon, indicating an instability in the theory.  We stabilize the theory by introducing a cosmological constant in the bulk and a pair of 3-branes, as in the Randall-Sundrum model.  The scalar glueball masses computed by lattice gauge theory are then described very well by just a few parameters.  Prospects for extending the model to other spins and parities and to finite temperature are considered.  

\end{abstract}

\maketitle
%\vfill

%%%%%%%%%%%%%
\section{Introduction}

The AdS/CFT correspondence has proven to be a useful mathematical tool for the analysis of certain strongly coupled gauge theories.
This correspondence establishes a connection between a d-dimensional super-Yang Mills (SYM) theory and a weakly coupled gravitational theory in d+1 dimensions \cite{maldacena, Gubser1998, Witten:1998}.  Perturbation theory can be used to solve SYM at small couping but it fails at large coupling, whereas the reverse is true in the gravity dual.  
QCD is a strongly-coupled gauge theory at hadronic scales, making it a prime candidate for the application of the gauge/gravity correspondence.
However, it is still not known whether a gravitational dual to QCD exists.  The so-called bottom-up approach assumes the existence of such a dual, modeling features of QCD by an effective five-dimensional gravity theory.  Linear confinement in QCD provides a scale that is embodied in an IR cut-off of the fifth dimension in the AdS/QCD model \cite{stephanov-katz-son, DaRold2005}; these are referred to as hard-wall models.  Soft-wall models use a dilaton field as an effective IR cutoff \cite{karch-katz-son-adsqcd}.  The simplest of these uses a dilaton quadratic in the fifth dimensional coordinate to provide linear radial Regge trajectories. Others modify the UV behavior of the dilaton to more accurately model the ground state masses as well \cite{gherghetta-kelley, bartz-pions, Colangelo2008, Cui2013}.

Typically, soft-wall models of QCD use parametrizations for the background dilaton and chiral fields that are not derived as the solution to any equations of motion. The problem is that there is always one more equation than there are independent fields, causing a severe self-consistency condition on the potential.  Previously, two of us expanded upon previous work \cite{Batell2008, Springer2010, Gursoy2008a, Gursoy2008b, Csaki2007, Li2013, Li2013a, He2013} to find a suitable potential for the background fields of a soft-wall AdS/QCD model \cite{3field}.  A glueball field was added and a potential was constructed which led to a dynamical dilaton and a good description of the mass spectra of vector, axial-vector, and pseudoscalar mesons.  A shortfall of that approach was that there was a special term in the potential, dependent on the dilaton field, that could only be determined numerically {\it a posteriori} and that did not appear to have a natural functional form.  It remains an open problem to find a well-defined, natural action that provides a set of background equations from which these fields can be derived.  A solution to that problem may suggest how it can be derived from a top-down approach. A well-defined action with good phenomenology is necessary to provide access to the thermal properties of the model through perturbation of the geometry \cite{Herzog2007, Bayona2008, Gursoy2008b}.

Our goal is to find a relatively simple analytic potential for Yang-Mills theory which is self-consistent, which satisifies the basic requirements of the gauge/gravity correspondence, and which reproduces the scalar glueball mass spectrum as computed in lattice gauge theory.  This would be useful not only for hadron phenomenology but might point towards a ten-dimensional string model from which it arises.  In addition, an analytic potential would greatly simplify the task of applying the model at finite temperature, such as the equation of state and the transport coefficients.  This task suggests using the tools of supersymmetry.  We will therefore follow Ref. \cite{Batell2008} which used two fields and a superpotential method.  In what follows one of the fields is the dilaton and the other is a scalar glueball.

We can illustrate the enduring interest in the relationship between Yang-Mills theory and the gauge/gravity correspondence with two papers separated by two decades.  Cs\'aki {\it et al.} \cite{Csaki} calculated the scalar glueball masses by solving supergravity wave equations in a black hole geometry and compared the results to the then-available lattice calculations.  Ballon-Bayona {\it et al.} \cite{Ballon} investigated effective holographic models for QCD and proposed semianalytic interpolations between the UV and the IR to obtain a spectrum for scalar and tensor glueballs consistent with current lattice QCD calculations.  In that approach the dilaton was identified as the glueball.  A UV cutoff has been a persistent issue, which we will address directly in our paper.

The outline of our paper is as follows.  In Sec. \ref{secReview} we review the results for scalar glueball masses computed in lattice gauge theory.  In Sec. \ref{setup} we give the basic setup for the five-dimensional action.  Some possible potentials which incorporate the proper behavior are given in Sec. \ref{potentials}.  A specific example is worked out in detail in Sec. \ref{logpot}.  We solve a Schr\"odinger equation for the glueball mass spectrum in Sec. \ref{gmass1} where it is found that the lowest state is a tachyon, which seems to be a generic result.  In order to stabilize the theory we consider the Randall-Sundrum model, which is briefly reviewed in Sec. \ref{RS}.  The dilaton and glueball fields are added to the Randall-Sundrum model in Sec. \ref{RSwith} and the glueball mass spectrum is recalculated in Sec. \ref{gmass2} with excellent results.  Our summary and conclusion is presented in Sec. \ref{conclude}.

%%%%%%%%%%%%%%%
\section{Lattice Results}
\label{secReview}

\begin{figure}
\center{\includegraphics[width=290pt]{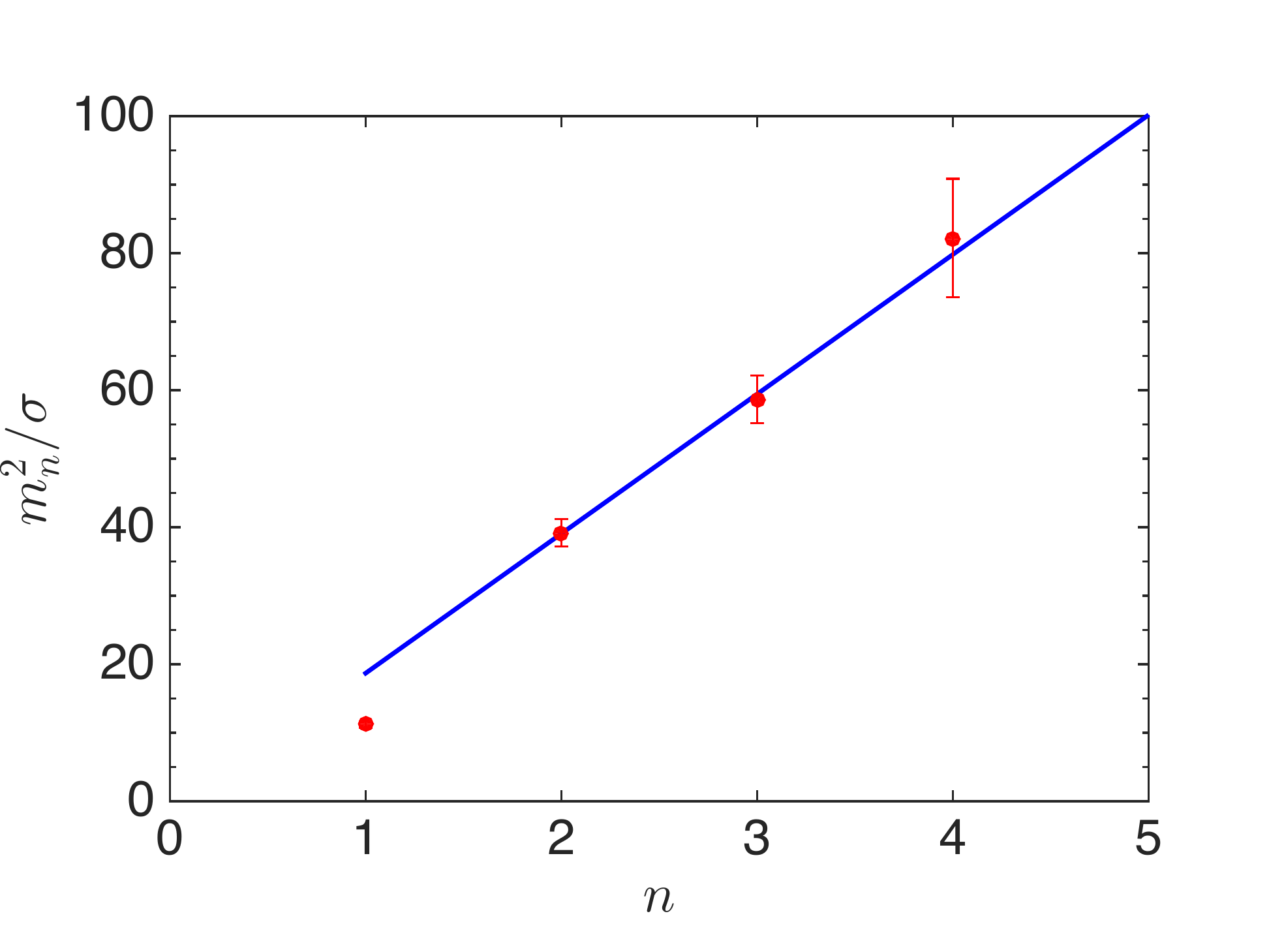}}
\caption{The scalar glueball mass spectrum as computed on the lattice in pure SU(3) gauge theory.  The linear fit includes $n \ge 2$.}
\label{lattice3}
\end{figure}

The $0^{++}$ glueball mass spectrum, as calculated by lattice gauge theory \cite{MeyerPhD, MeyerTeper2005}, is shown in Fig. \ref{lattice3}.  The radial quantum number is $n$ and the masses are expressed in terms of the string tension $\sigma = (440 \, {\rm MeV})^2$.  The error bars include statistical uncertainties only.  It is apparent that the spectrum is linear at large $n$.  A linear fit to the states $n = 2,3,4$ of the form $m_n^2/\sigma = a n + b$ results in $a = 20.37 \pm 1.142$ and $b = - 1.666 \pm 2.713$.  For later use we express this as
\be
m_n^2 = 8 \lambda (n - 0.082 \pm 0.133)
\label{Lspectrum3}
\ee
with $\sqrt{\lambda} = 702$ MeV.  Here we are not interested in the small uncertainty in the slope, only the intercept which is consistent with zero.  

\begin{figure}
\center{\includegraphics[width=290pt]{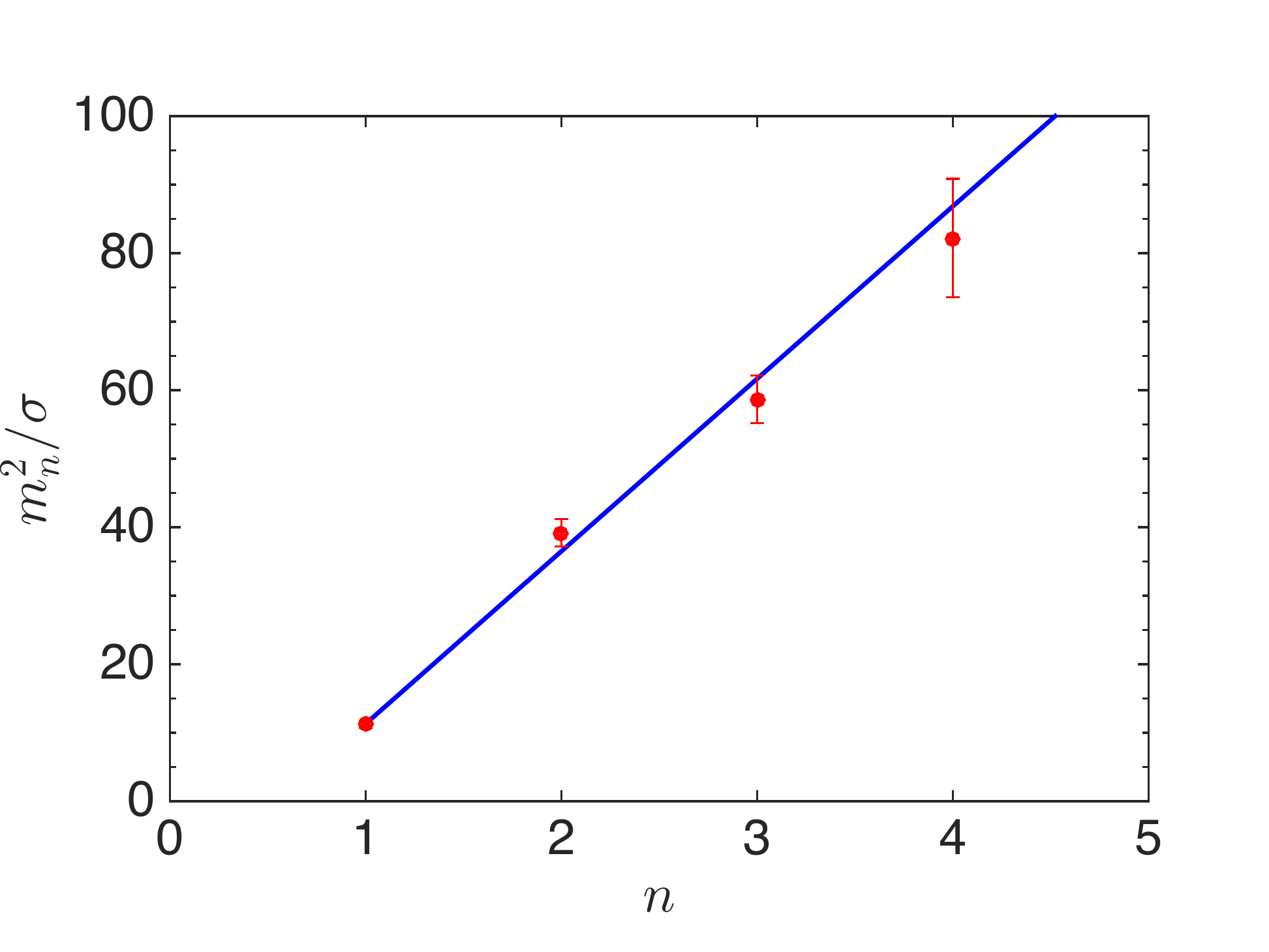}}
\caption{The scalar glueball mass spectrum as computed on the lattice in pure SU(3) gauge theory.  The linear fit includes $n \ge 1$.}
\label{lattice4}
\end{figure}

We can also make a linear fit to all states including $n=1$ as shown in Fig. \ref{lattice4}.  The form $m_n^2/\sigma = a n + b$ results in $a = 25.178 \pm 1.483$ and $b = - 13.859 \pm 1.703$.  Then
\be
m_n^2 = 8 \lambda (n - 0.550 \pm 0.075)
\label{Lspectrum4}
\ee
with $\sqrt{\lambda} = 781$ MeV.  In this fit the intercept is clearly negative and the fit is not nearly as good.

%%%%%%%%%%%%%%%
\section{Setup}
\label{setup}

We assume that the four-dimensional strongly coupled field theory can be modeled by the following five-dimensional action, written in the string frame:
\be
\cS = \frac{1}{16\pi G_5} \int d^5x \sqrt{g} e^{-2\Phi}  \Bigg( R+4 g^{MN} \partial_M\Phi\partial_N\Phi  -  \thalf g^{MN} \partial_MG\partial_NG - V(\Phi, G) \Bigg) \, .
\label{eqStringAction}
\ee
Here $\Phi$ is the dilaton and $G$ is the glueball field; these fields are dimensionless.  The metric is pure anti-de Sitter (AdS), $g_{MN}=(l/z)^2\eta_{MN}$ with AdS curvature $R = -30/l^2$ and $g = - \det g_{MN}$.

It is easier to search for the background fields in the Einstein frame, which is obtained from the string frame via the conformal transformation
\be
g_{MN}=e^{4\Phi/3}\tilde{g}_{MN} \, .
\ee
where the tilde distinguishes the two frames.  Rescaling the dilaton by $\phi=\sqrt{8/3}\Phi$ puts the vacuum action in canonical form
\be
\cS_E=\frac{1}{16\pi G_5} \int d^5x \sqrt{\tilde{g}}\left(\tilde{R}-\thalf \tilde{g}^{MN} \partial_M\phi\partial_N\phi -\thalf \tilde{g}^{MN} \partial_M G \partial_N G - 
\tilde{V}(\phi,G)\right) \, ,
\label{eq:Einstein}
\ee
with $\tilde{V}=e^{4\Phi/3}V = e^{2\phi/\sqrt{6}}V$.

To apply the superpotential method \cite{Skenderis1999,DeWolfe2000} requires switching from $(x^{\mu},z)$ to $(x^{\mu},y)$ coordinates such that
\be
ds^2 = e^{-2A(y)}dx^2 + dy^2 \, .
\ee
We look for a background solution with $\phi = \phi(y)$ and $G = G(y)$.  The equations of motion are
\be
3 A'' - 6 A'^2 = \oneqt \phi'^2 + \oneqt G'^2 + \thalf\tilde{V} \, ,
\ee
\be
6 A'^2 = \oneqt \phi'^2 + \oneqt G'^2 - \thalf\tilde{V} \, ,
\ee
\be
\phi'' - 4 A' \phi' =  \frac{\partial \tilde{V}}{\partial \phi}\, ,
\ee
\be
G'' - 4 A' G' =  \frac{\partial \tilde{V}}{\partial G}\, .
\ee
Here a prime means differentiation with respect to $y$.  Since there is one more equation than there are independent fields, these equations would not have a solution for an arbitrarily chosen potential $\tilde{V}(\phi,G)$.  To solve these equations we convert them to a set of first order differential equations by use of a ``superpotential" $W(\phi,G)$ such that
\be
A' = W \, ,
\ee
\be
\phi' = 6 \frac{\partial W}{\partial \phi}\, ,
\ee
\be
G' = 6 \frac{\partial W}{\partial G}\, .
\ee
Then the original set of equations is satisifed when
\be
\tilde{V} = 18 \left[ \left( \frac{\partial W}{\partial \phi} \right)^2 + \left( \frac{\partial W}{\partial G} \right)^2 \right] - 12 W^2 \, .
\ee
The challenge is to find a function $W(\phi,G)$ which gives the desired behavior of the fields.  But it must be emphasized that any appropriately smooth function $W$ will allow for all the original differential equations to be satisfied, which is our goal.

As discussed above, good hadron phenomenology suggests that $A$, $\phi$, $y$ and $z$ are related in the following ways.
\be
{\rm e}^{-A} = \frac{l}{z}\,{\rm e}^{-a\phi}
\ee
\be
dy =  \frac{l}{z}\,{\rm e}^{-a\phi} \, dz
\ee
Here $a$ is an arbitrary constant.  From the first of these equations we clearly have $A = \ln(z/l) + a \phi$.  Now the differential equation for $A$ becomes
\be
A' = \frac{1}{z}\frac{dz}{dy} + a \phi' = W \, .
\ee
Using the relationship between $y$ and $z$ and the differential equation for $\phi$ we get a differential equation for $W$.
\be
\frac{\partial W}{\partial \phi} = \frac{1}{6a} \left( W - \frac{1}{l} \, {\rm e}^{a\phi} \right)
\ee
As we know from previous studies \cite{Springer2010} the constant $a=1/\rt6$ in order to have linear radial Regge trajectories; we shall henceforth fix it at that value.  Then the most general solution to the differential equation for $W$ is
\be
W(\phi,G) =\frac{1}{l} \left[ U(G) + 1 - \frac{\phi}{\rt6} \right] e^{\phi/\rt6} \, .
\label{W1}
\ee
Here $U$ is any function of $G$; the +1 is added for future convenience.  The resulting differential equations can be written simply as
\be
z \frac{dG}{dz} = 6 \frac{dU}{dG} \, ,
\label{Tdiffeq}
\ee
\be
\frac{d}{dz}(z \phi) = \rt6 \, U \, .
\label{phidiffeq}
\ee
Alternatively the first equation can be inserted in the second to obtain
\be
6 \frac{dU}{dG} \frac{d\phi}{dG}+ \phi = \rt6 \, U \, .
\ee
This is a remarkable simplification from previous studies of phenomenologically relevant AdS-inspired approaches.

The potential resulting from this analysis is
\be
l^2 V(\phi,G) = -12 + 4 \rt6 \phi - \tthalf \phi^2 + 3 \rt6 \phi U -24 U -9 U^2 + 18 \left(\frac{dU}{dG}\right)^2 \, .
\label{Vexact}
\ee
One might wonder about the term linear in $\phi$, but really it should be viewed in the Einstein frame.  Assuming that $U$ has no constant terms
\be
l^2 \tilde{V} = -12 + \textstyle{\frac{5}{2}} \phi^2 + \cdot\cdot\cdot \, .
\ee
The first term is just the usual negative cosmological constant, while the second term suggests that the mass of the dilaton is $m_{\phi}^2 l^2 = 5$.  If one wants to relate this to an operator it would have dimension 5.  However, as pointed out in \cite{Springer2010}, the AdS/CFT correspondence between dimension of the operator and mass of the field does not generally apply to the dilaton field for the reasons presented there.

%%%%%%%%%%%%%%%%%%%%%
\section{Possible Potentials}
\label{potentials}

The AdS/CFT dictionary sets the mass for the fields according to the dimension $\Delta$ of the dual operator,
\be
m^2l^2=\Delta(\Delta-4) \,,
\ee
where $l$ is related to the AdS curvature.  The glueball mass is 0 since the dimension of the gluon condensate is 4.  The dilaton mass is undetermined and is not connected to the dimension of the corresponding operator, as discussed in \cite{Springer2010}.

In the UV limit, $z \rightarrow 0$, the glueball field is proportional to $z^4$.  In the IR limit, $z \rightarrow \infty$, the glueball field must be proportional to $z$ in order that the dilaton field be proportional to $z^2$ so that the mesons have linear radial Regge trajectories.  As a consequence $U \rightarrow {\textstyle{\frac{1}{12}}} G^2$ in the IR limit.  In the UV limit  $U \rightarrow \oneth G^2$.

One possible potential which yields the correct UV and IR behavior is
\be
U = {\textstyle{\frac{1}{12}}} G^2 \left[ 1 + 3 {\rm e}^{-\gamma G^2} \right] \,.
\ee
Here the only free parameter is $\gamma$ which controls the transitional behavior of $G$ between the IR and the UV.

A second possible potential is
\be
U = {\textstyle{\frac{1}{12}}} G^2 \left[ 4 - 3 \tanh{\gamma G^2} \right] \,.
\ee
The free parameter $\gamma$ controls the transitional behavior.

A third possible potential is
\be
U = {\textstyle{\frac{1}{12}}} G^2 + \frac{1}{4\gamma} \ln \left( 1 + \gamma G^2 \right) \,.
\ee
Again the transitional behavior is fixed by the numerical value of $\gamma$.

%%%%%%%%%%%%%%%%%%%
\section{Log Potential}
\label{logpot}

\begin{figure}[th]
\center{\includegraphics[width=290pt]{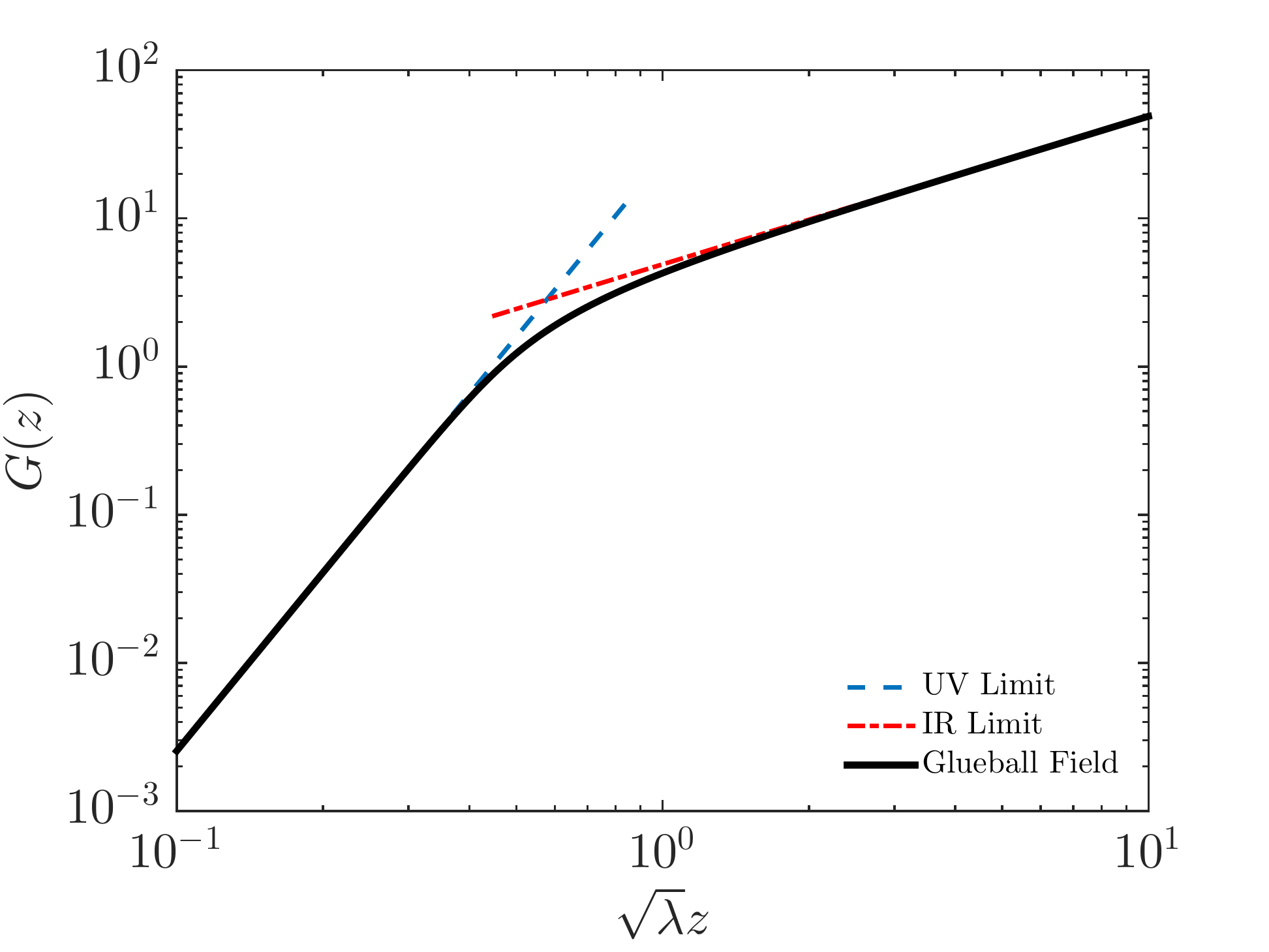}}
\caption{The glueball field as a function of the dimensionless variable $\sqrt{\lambda}z$.  Plotted this way it only depends on the parameter $\gamma$ which here is chosen to be 1/2.}
\label{Glueball700}
\end{figure}

Let us focus on the logarithmic potential because we can find some analytical results straightaway.  The differential equation for $G$ 
\be
z \frac{dG}{dz} = 6 \frac{dU}{dG} = \frac{\gamma G^2 + 4}{\gamma G^2 + 1} \, G
\ee
has the solution
\be
G^2 \left(\gamma G^2 + 4\right)^3 = \left(\frac{z}{z_0}\right)^8 \, .
\label{Glogsol}
\ee
Here $z_0$ is a constant of integration.  In the large $z$ limit the solution is
\be
\gamma G^2 = \gamma^{1/4} \left(\frac{z}{z_0}\right)^2 - 3 + \cdot\cdot\cdot \, .
\ee
In the small $z$ limit the solution is
\be
G = \frac{1}{8} \left(\frac{z}{z_0}\right)^4 - 3 \gamma \left(\frac{z}{2z_0}\right)^{12} + \cdot\cdot\cdot \, .
\ee
The field is plotted as a function of $z$ in Fig. \ref{Glueball700} (see Eq. (\ref{z0lambda}) for the relationship among $z_0$, $\gamma$ and $\lambda$). The power-law behavior in the UV and IR limits is evident.

\begin{figure}[th]
\center{\includegraphics[width=290pt]{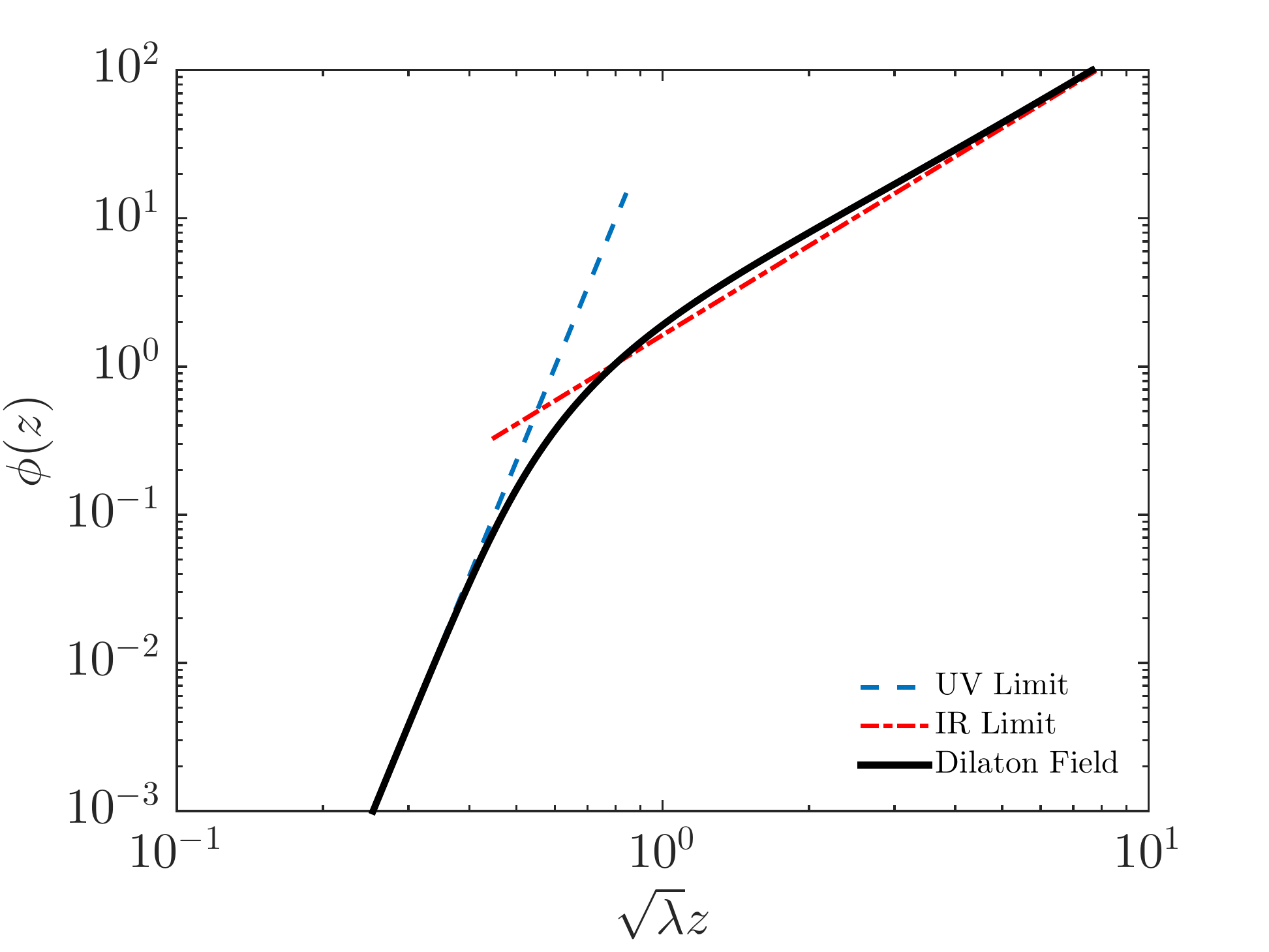}}
\caption{The dilaton field as a function of the dimensionless variable $\sqrt{\lambda}z$.  Plotted this way it only depends on the parameter $\gamma$ which here is chosen to be 1/2.}
\label{Dilaton700}
\end{figure}

We have not found a solution $\phi(G)$ in terms of elementary functions.  The solution can be expressed in terms of an integral with the proper boundary conditions as
\be
\phi(G) = \rt6 \left\{ U(G) - \frac{z_0}{z(G)} \int_0^G \frac{z(G')}{z_0} \frac{dU}{dG'} dG' \right\}
\ee
where
\be
\frac{dU}{dG} = \frac{G}{6} \, \frac{4 + \gamma G^2}{1 + \gamma G^2}
\ee
and $z/z_0$ is given by Eq. (\ref{Glogsol}).  It is straightforward to find the limits as $G \rightarrow \infty$
\ba
\phi &=& \rt6 \left[ \frac{1}{36} G^2 + \frac{1}{4\gamma} \ln \gamma G^2 - \frac{2}{3\gamma} + \cdot\cdot\cdot \right] \nonumber \\ 
&=& \rt6 \left[ \frac{1}{36 \gamma^{3/4}} \left(\frac{z}{z_0}\right)^2 + \frac{1}{4\gamma} \ln\left(\frac{z}{z_0}\right)^2 + \frac{1}{16\gamma} \ln\gamma - \frac{3}{4\gamma}
+ \cdot\cdot\cdot \right] 
\ea
and as $G \rightarrow 0$
\ba 
\phi &=& \frac{\rt6}{9} \left[ \frac{1}{3} G^2 + \frac{7}{136} \gamma G^4 + \cdot\cdot\cdot \right] \nonumber \\
&=& \frac{2\rt6}{9} \left[ \frac{2}{3} \left(\frac{z}{2z_0}\right)^8 - \frac{27 \gamma}{17} \left(\frac{z}{2z_0}\right)^{16} \cdot\cdot\cdot \right] \, .
\ea
The requirement of linear confinement requires a solution in the large $z$ limit of the  form
\be
\phi = \sqrt{\frac{8}{3}}\lambda z^2 \, .
\ee
Using the above limiting forms we find that the constant of integration in the differential equation for $G(z)$ is
\be
\frac{1}{z_0^2} = 24 \gamma^{3/4} \lambda \, .
\label{z0lambda}
\ee
This also determines the magnitude of the gluon condensate, defined by $G = G_4 z^4$, to be
\be
G_4 = 72 \gamma^{3/2} \lambda^2 \, .
\ee
Only $\gamma$ remains as an undetermined parameter.  The field is plotted as a function of $z$ in Fig. \ref{Dilaton700}. The power-law behavior in the UV and IR limits is evident.

%%%%%%%%%%%%%%%
\section{Glueball Mass Spectrum}
\label{gmass1}

Suppose that the metric has the form
\be
ds^2 = b^2(z) \left( -dt^2 + d{\bf x}^2 + dz^2 \right)
\label{S1}
\ee
and that the action has the canonical form of Eq. (\ref{eq:Einstein}). The equation of motion which determines the mass spectrum for the field $G = G(z) + \Delta G(z,x)$ is
\be
-b^3 \eta^{\mu\nu}\partial_{\mu}\partial_{\nu} \Delta G - \partial_z(b^3 \partial_z \Delta G) + b^5 \frac{\partial^2\tilde{V}}{\partial G^2} \Delta G = 0 \,.
\label{S2}
\ee
Here $\Delta G(z,x) = \Delta G_n(z) {\rm e}^{i p \cdot x}$ so that $\eta^{\mu\nu}\partial_{\mu}\partial_{\nu} \rightarrow m^2$.  This leads to the eigenvalue equation
\be
-\partial_z^2 \Delta G_n(z) - b^{-3} \left( \partial_z b^3 \right) \partial_z \Delta G_n(z) + b^2 \frac{\partial^2\tilde{V}}{\partial G^2} \Delta G_n(z) = m_n^2 \Delta G_n(z) \,.
\label{S3}
\ee
It can be put in the form of the Schr\"odinger equation by the transformation $\Delta G_n = {\rm e}^{\omega_s/2} H_n$ with $\omega_s = - 3 \ln b$.  Hence
\be
- \ddot{H}_{n} + \left[ \oneqt \dot{\omega_s}^2-\thalf \ddot{\omega}_s +  \frac{l^2}{z^2}  \frac{\partial^2 V}{\partial G^2} \right]  H_{n} = m_{n}^2 H_{n}
\ee
where a dot means differentiation with respect to $z$.

In the present case $\omega_s = \sqrt{\tthalf} \phi + 3 \ln z$.  From the equations of motion one can write
\ba
z \dot{\phi} &=& \rt6 U - \phi \nonumber \\
z^2 \ddot{\phi} &=& 2 \phi - 2 \rt6 U + 6\rt6 \left( \frac{\partial U}{\partial G} \right)^2 \, .
\label{dots}
\ea
These can be used in
\be
\dot{\omega}_s = \frac{1}{z} \left( 3 + \sqrt{\frac{3}{2}} z \dot{\phi} \right)
\ee
and in
\be
\ddot{\omega}_s = \frac{1}{z^2} \left( -3 + \sqrt{\frac{3}{2}} z^2 \ddot{\phi} \right)
\ee
without having to numerically calculate the first or second derivatives of $\phi$ with respect to $z$.  This combines to give
\be
\oneqt \dot{\omega_s}^2-\thalf \ddot{\omega}_s = \frac{1}{z^2} \left[ \frac{15}{4} -\frac{5}{4} \rt6 \phi + \frac{15}{2} U + \frac{9}{4} U^2   -\frac{3}{4} \rt6 \phi U
+ \frac{3}{8} \phi^2 - 9 \left(\frac{dU}{dG}\right)^2 \right] \,.
\ee
The potential in Eq. (\ref{Vexact}) gives
\be
l^2 \frac{\partial^2 V}{\partial G^2} = 3\sqrt{6}\phi\frac{d^2 U}{dG^2} - 24\frac{d^2 U}{dG^2} - 18U\frac{d^2 U}{dG^2} - 18\left(\frac{dU}{dG}\right)^2 + 36\left(\frac{d^2U}{dG^2}\right)^2 + 36\frac{dU}{dG}\frac{d^3U}{dG^3} 
\ee
which does not specify the function $U(G)$.

For the specific example of the logarithmic form of $U$ we find
\ba
\frac{dU}{dG} &=& \frac{G}{6} \left[ \frac{\gamma G^2 + 4}{\gamma G^2 + 1} \right] \\
\frac{d^2 U}{dG^2} &=& \frac{1}{6}\left[\frac{(\gamma G^2)^2-\gamma G^2 + 4}{(\gamma G^2 + 1)^2}\right] \\
\frac{d^3 U}{dG^3} &=& \frac{ \gamma G (\gamma G^2 - 3)}{(\gamma G^2+1)^3} \,.
\ea
The Schr\"odinger equation can be put in dimensionless form by changing to the dimensionless variable $x = \sqrt{2\lambda} z$.  Then
\be
- \frac{d^2H_{n}}{dx^2} + V_H\left(\phi(x), G(x) \right)  H_{n} = \frac{m_{n}^2}{2\lambda} H_{n} \,.
\label{Hdiffeq3}
\ee
where
\be
V_H = \frac{1}{x^2} \left[\frac{15}{4} -\frac{5}{4} \rt6 \phi + \frac{15}{2} U + \frac{9}{4} U^2   -\frac{3}{4} \rt6 \phi U
+ \frac{3}{8} \phi^2 - 9 \left(\frac{dU}{dG}\right)^2 + l^2 \frac{\partial^2 V}{\partial G^2} \right] \,.
\label{pot1}
\ee
It is straightforward to compute the limits of $V_H$ in the UV and IR limits.  In the IR $V_H \rightarrow x^2$ and in the UV $V_H \rightarrow 15/4x^2$.  These are independent of the parameter $\gamma$. 

\begin{figure}[th]
\center{\includegraphics[width=290pt]{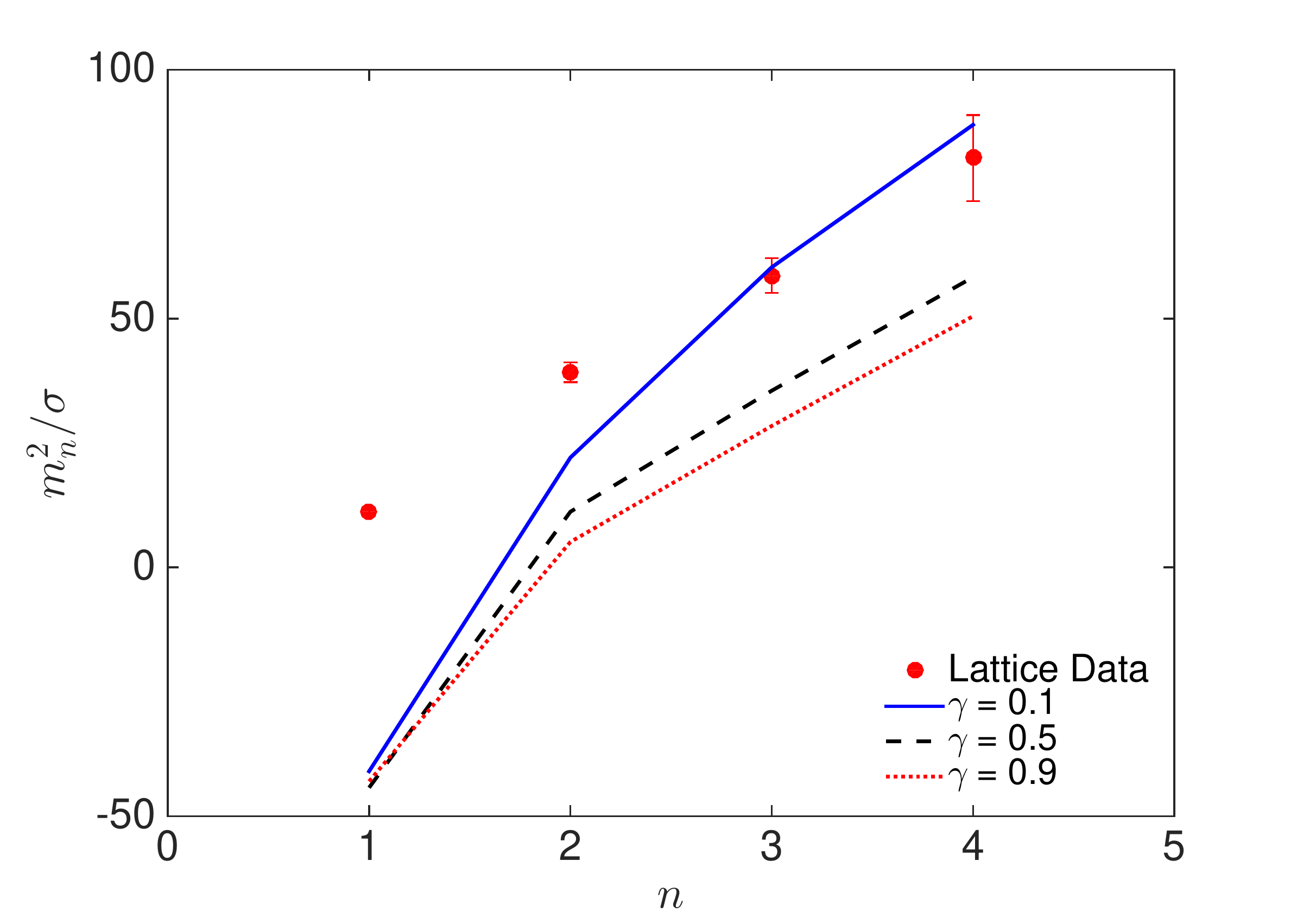}}
\caption{The mass spectra with $\sqrt{\lambda} = 700$ MeV for several values of the parameter $\gamma$.}
\label{spectra_log1}
\end{figure}

Now the parameter $\gamma$ should be used to obtain the best fit to the glueball mass spectrum.  Some results are plotted in Fig. \ref{spectra_log1} for a representative range of $\gamma$.  It shows the existence of a tachyon when $n=1$.  Very similar results are obtained with the other forms of $U$ discussed in Sec. \ref{potentials}.

In fact this can be anticipated analytically by consideration of the limit $\gamma >> 1$.  Then the mass spectrum is obtained from 
\be
- \frac{d^2H_{n'}}{dx^2} + \left[ x^2 - 6  + \frac{3}{4x^2}   \right]  H_{n'} = \frac{m_{n'}^2}{2\lambda} H_{n'} \,.
\label{pot2}
\ee
This may be compared to the radial equation for the harmonic oscillator in the form
\be
- \frac{d^2H_{n'}}{dx^2} + \left[ x^2 + \frac{l^2-1/4}{x^2}  \right]  H_{n'} = \frac{2 E_{n'}}{\hbar \omega} H_{n'}
\label{2de}
\ee
with radial quantum number $n' = 0, 1, 2....$, $l = \thalf, \tthalf, ...$ and energy eigenvalue
\be
E_{n'} = \hbar \omega \left[ 2n' + l + 1 \right] \, .
\label{2dE}
\ee
Switching to the radial quantum number $n=n'+1$ to be consistent with the notation used earlier results in 
\be
m_n^2 = 8\lambda \left[ n  - \frac{3}{2} \right] \,.
\label{m2n}
\ee
This has exactly one tachyonic state.

%%%%%%%%%%%%%%%%%%%%%
\section{Randall-Sundrum Model}
\label{RS}

There is a tachyon in the versions of the AdS model considered above which indicates an instability.  In an attempt at stabilization we turn to the Randall-Sundrum model \cite{RS1}.  The goal of Ref. \cite{RS1} was to obtain a more natural explanation of the hierarchy between the weak scale and the fundamental scale of gravity.  Here we have more mundane considerations.

The Randall-Sundrum model introduces a cosmological constant $\Lambda$ in the bulk and two 3-branes, one located at $y=0$ and the other at $y=L$.  The 3-branes have constant energy densities $V_0$ and $V_L$, respectively.  There is an assumed symmetry $y \rightarrow -y$ so that only positive values of $y$ are needed.  The action is
\be
\cS = 2M^3 \int d^4x dy \sqrt{g} \left( R + \frac{\Lambda}{2M^3} \right) - \int d^4x \sqrt{g(y=0)} V_0 - \int d^4x \sqrt{g(y=L)} V_L \,.
\label{eq:RS}
\ee
Here $M$ is the five-dimensional gravity scale ($M^3 = 1/32\pi G_5$).  The metric has the form
\be
ds^2 = {\rm e}^{-2A(y)} (-dt^2 + d{\bf x}^2) + dy^2 \,,
\ee
with indices running from 0 through 4, 4 representing the fifth dimension $y$.  Note that we have dropped the tildes in this section even though we are working in the Einstein frame.  The sign in front of $\Lambda$ in the action is chosen so that in this model it turns out positive.  The formulas for the components of the Ricci tensor are given in the Appendix.  For this metric they are
\be
R_{00} = -(A'' - 4A'^2) {\rm e}^{-2A} = -R_{ii}
\ee
with no sum on $i$ = 1,2,3 and 
\be
R_{yy} = 4A'' - 4 A'^2  \,.
\ee
The scalar curvature is
\be
R = 8A'' -  20 A'^2 \,.
\ee

Einstein's equations give the following pair of scalar equations.
\be
A'^2 = \frac{\Lambda}{24M^3} \equiv k^2
\label{A1}
\ee
\be
A'' - 2 A'^2 = - \frac{\Lambda}{12M^3} + \frac{V_0}{12M^3} \delta(y) + \frac{V_L}{12M^3} \delta(y-L)
\label{A2}
\ee
The solution to Eq. (\ref{A1}) is simply
\be
A(y) = k |y| \theta(L-|y|) 
\ee
when making using of the symmetry $y \rightarrow -y$.  Also from Eq. (\ref{A1}) we have
\be
A' = k \left[ \theta(y) \theta(L-y) - \theta(-y) \theta(L+y) \right] \,.
\ee
Differentiating this expression, using the symmetry, and comparing to Eq. (\ref{A2}), we find the requirements that $V_0 = \sqrt{24M^3\Lambda}$ and $V_L = -\sqrt{24M^3\Lambda}$.  These requirements are equivalent to demanding that the four-dimensional cosmological constant be zero.  For this, we need
\ba
\int_{-L}^L  dy \sqrt{g(y)} &=& \frac{1}{2k} \left( 1 - {\rm e}^{-4kL} \right) \,, \nonumber \\
\int_{-L}^L  dy \sqrt{g(y=0)} &=& \frac{1}{2k} \,, \nonumber \\
\int_{-L}^L  dy \sqrt{g(y=L)} &=& \frac{1}{2k} {\rm e}^{-4kL} \,, \nonumber \\
\int_{-L}^L  dy \sqrt{g(y)} \; R(y) &=& - 6 k \left( 1 - {\rm e}^{-4kL}  \right) \,.
\ea
Then the four-dimensional action is
\be
\int d^4x \left[ 12 M^3 k \left( 1 - {\rm e}^{-4kL} \right) + \frac{\Lambda}{2k} \left( 1 - {\rm e}^{-4kL} \right) - \left( V_0 + V_L {\rm e}^{-4kL} \right) \right]
\label{4dA}
\ee
which is zero if $V_0$ and $V_L$ take on the values mentioned above. 

%%%%%%%%%%%%%%%%%%%%%
\section{Randall-Sundrum Model with Scalar Fields}
\label{RSwith}

Now we add the scalar fields $\phi$ and $G$ as in Sec. III.  We assume that $V_L = - V_0$ but $V_0$, $\Lambda$, and $M$ are unrelated; $k$ is still defined as $\sqrt{\Lambda/24M^3}$.  The equations of motion are
\be
3 A'' - 6 A'^2 + \frac{\Lambda}{4M^3} = \oneqt \phi'^2 + \oneqt G'^2 + \thalf\tilde{V} + \frac{V_0}{4M^3} [ \delta(y) - \delta(y-L) ] \, ,
\label{Adprime}
\ee
\be
6 A'^2 - \frac{\Lambda}{4M^3} = \oneqt \phi'^2 + \oneqt G'^2 - \thalf\tilde{V} \, ,
\label{Aprime}
\ee
\be
\phi'' - 4 A' \phi' =  \frac{\partial \tilde{V}}{\partial \phi}\, ,
\ee
\be
G'' - 4 A' G' =  \frac{\partial \tilde{V}}{\partial G}\, .
\ee
Based on previous results we look for a solution of the following form:
\be
A' = W + k \left[ \theta(y) \theta(L-y) - \theta(-y) \theta(L+y) \right] \, ,
\ee
\be
\phi' = 6 \frac{\partial W}{\partial \phi}\, ,
\ee
\be
G' = 6 \frac{\partial W}{\partial G}\, .
\ee
Then the original set of equations are satisfied when
\be
\tilde{V} = 18 \left[ \left( \frac{\partial W}{\partial \phi} \right)^2 + \left( \frac{\partial W}{\partial G} \right)^2 \right] - 12 W^2 
- 24 k \left[ \theta(y) \theta(L-y) - \theta(-y) \theta(L+y) \right] W \, ,
\ee
and $V_0 = \sqrt{24M^3\Lambda}$, as before.  

\subsection{Change of coordinate}

Reference \cite{RS2} studied tensor fluctuations in the absence of scalar fields.  The authors found it convenient to make the change of variables from $y$ to $z$ via
\be
z = {\rm sign}(y) \left[ {\rm e}^{k|y|} -1 \right]/k \,.
\ee
With the inclusion of scalar fields it suggests that we choose
\be
A(y) = \ln\left(k|z|+1\right) + a \phi(|z|) \,,
\ee
valid for both positive and negative values of $z$ including $z=0$.  Then $y=0$ corresponds to $z=0$.  Along with this we assume that
\be
dy = \frac{{\rm e}^{-a\phi(|z|)}}{k|z|+1}  dz \,.
\ee
This gives the metric
\be
ds^2 = \frac{{\rm e}^{-2a\phi(|z|)}}{(k|z|+1)^2}  \left( -dt^2 + d{\bf x}^2 + dz^2 \right) \,.
\ee
The equation for $A$ is now
\be
A' = \frac{{\rm sign}(z)}{|z|+l}\frac{dz}{dy} + a \phi' = W + k \, {\rm sign}(z) \, .
\ee
Unless otherwise stated we shall assume that $L$ becomes arbitrarily large so that we may ignore any singularities at $y = \pm L$ in the equations of motion.

Using the equations above we can readily find a differential equation which determines the $\phi$ dependence of $W$.
\be
6a \frac{\partial W}{\partial \phi} = W + k \, {\rm sign}(z)  \left( 1 - {\rm e}^{a\phi} \right)
\ee
The solution is
\be
W = k \, {\rm sign}(z) \left\{ \left[ U(G) + 1 - a \phi \right] {\rm e}^{a\phi} -1 \right\} \,.
\label{W2}
\ee
Compare this to Eq. (\ref{W1}).  Therefore we can write the potential $\tilde{V}$ as
\be
\tilde{V} = 18 \left[ \left( \frac{\partial W}{\partial \phi} \right)^2 + \left( \frac{\partial W}{\partial G} \right)^2 \right] - 12 W^2 
- 24 k |W| \, .
\ee
The potential $V$ for non-negative values of $z$ is nearly the same as before.
\be
V(\phi,G)/k^2 = -12\left( 1 - {\rm e}^{-2a\phi} \right) + 4 \rt6 \phi - \tthalf \phi^2 + 3 \rt6 \phi U -24 U -9 U^2 + 18 \left(\frac{dU}{dG}\right)^2
\label{VexactRS}
\ee 
The only difference is the term proportional to ${\rm e}^{-2a\phi}$ which arises on account of the brane at $z=0$.

In terms of the $z$ coordinate the equations for the fields are now ($z \ge 0$)
\ba
(z+l) \frac{dG}{dz} &=& 6 \, \frac{dU}{dG} \\
(z+l) \frac{d\phi}{dz} + \phi &=& \rt6 \, U \,.
\ea
The $l$ acts as a UV cutoff.  Compare these to Eqs. (\ref{Tdiffeq}) and (\ref{phidiffeq}).  The solutions with the 3-brane, $G(z,l)$ and $\phi(z,l)$, can be obtained from the solutions without the 3-brane, $G(z,0)$ and $\phi(z,0)$, simply by shifting the coordinate, specifically $G(z,l) = G(z+l,0)$ and $\phi(z,l) = \phi(z+l,0)$.

\subsection{Vacuum energy density in four dimensions with scalar fields}

Let us calculate the vacuum energy density in four dimensions with the inclusion of the fields $\phi$ and $G$.  The action is now a combination of Eqs. (\ref{eq:Einstein}) and (\ref{eq:RS}) with
\be
\tilde{\cal L} = -\thalf \tilde{g}^{MN} \partial_M\phi\partial_N\phi -\thalf \tilde{g}^{MN} \partial_M G \partial_N G - \tilde{V}(\phi,G)
\ee
and
\be
\tilde{R} = 8A'' - 20 A'^2 \,.
\ee
From Eqs. (\ref{Adprime}) and (\ref{Aprime}) we may infer that
\be
\tilde{\cal L} = -6A'' + 12 A'^2 - \frac{\Lambda}{2M^3} +  \frac{V_0}{2M^3} [ \delta(y) - \delta(y-L) ]
\ee
when the solutions are plugged back into the Lagrangian.  Then the action is
\be
\cS_E = 4M^3 \int d^4x \int dy \, {\rm e}^{-4A} (A'' - 4 A'^2) \,.
\ee
Again using the same equations we have
\be
A'' - 4 A'^2 = \oneth \tilde{V} - 4k^2 + \frac{V_0}{12M^3} \left[ \delta(y) - \delta(y-L)\right] \,.
\ee
The last pair of terms integrates to zero leaving
\be
\cS_E = \frac{4}{3}M^3 \int d^4x \int dy \, {\rm e}^{-4A} \tilde{V} \,.
\ee

Is this action finite?  Switching from $y$ to $z$ involves the change
\be
dy \, {\rm e}^{-4A} = dz \frac{{\rm e}^{-5a\phi}}{(k|z|+1)^5} \, .
\ee
The action is thus proportional to
\be
\int_0^{\infty} \frac{dz}{(kz+1)^5} \, {\rm e}^{-3a\phi} \,V
\ee
In the IR, $V \rightarrow -6\phi^2(z)$, while in the UV, $V \rightarrow$ constant.  Hence the action is finite and so is the four-dimensional vacuum energy density.  This is due to the UV cutoff $k$ in the Randall-Sundrum model.

%%%%%%%%%%%%%%%%%%
\section{Glueball Mass Spectrum Revisited}
\label{gmass2}

In this section we calculate the glueball mass spectrum with the Randall-Sundrum modified action described above.  From Eqs. (\ref{S1})-(\ref{S3}) we find the Schr\"odinger equation 
\be
- \ddot{H}_{n} + \left[ \oneqt \dot{\omega_s}^2-\thalf \ddot{\omega}_s + \frac{1}{(k|z|+1)^2}  \frac{\partial^2 V}{\partial G^2} \right]  H_{n} = m_{n}^2 H_{n}
\ee
where
\ba
\omega_s &=& 3 \ln\left(k|z|+1\right) + \sqrt{\frac{3}{2}} \phi(|z|) \nonumber \\
\dot{\omega}_s &=& \left\{ \frac{3k}{k|z|+1}  + \sqrt{\frac{3}{2}} \frac{d}{d|z|} \phi(|z|) \right\} {\rm sign}(z) \nonumber \\
\ddot{\omega}_s &=& - \frac{3k^2}{(k|z|+1)^2} + \sqrt{\frac{3}{2}} \ddot{\phi} + 6 k \delta(z) \,.
\ea
Restricting $z$ to positive values only results in the following equation.
\be
- \ddot{H}_{n} + \left[ \frac{15}{4} \frac{k^2}{(kz+1)^2} + \frac{3}{2}\sqrt{\frac{3}{2}} \frac{k\dot{\phi}}{kz+1} + \frac{3}{8} \dot{\phi}^2 
- \frac{1}{2} \sqrt{\frac{3}{2}} \ddot{\phi} -\frac{3}{2} k \delta(z)
 + \frac{1}{(kz+1)^2}  \frac{\partial^2 V}{\partial G^2} \right]  H_{n} = m_{n}^2 H_{n}
\ee
Note the appearance of an attractive $\delta$-function potential.  Its coefficient is 3/2 instead of 3 because we are restricting the space to non-negative values of $z$.  We can express $\dot{\phi}$ and $\ddot{\phi}$ in terms of the background fields and $z$ as before using Eq. (\ref{dots})
\be
- \ddot{H}_{n} + V_H H_n = m_{n}^2 H_{n}
\ee
where now
\be
V_H = \frac{k^2}{(kz+1)^2} \left[ \frac{15}{4} + \frac{15}{2}  (U - a\phi) + \frac{9}{4} (U - a\phi)^2
 - 9 \left( \frac{dU}{dG}\right)^2  + \frac{1}{k^2}  \frac{\partial^2 V}{\partial G^2} \right]
-\frac{3}{2} k \delta(z) \,.
\ee
The Schr\"odinger equation may be solved for positive values of $z$ by neglecting the Dirac $\delta$-function but imposing the boundary condition 
$\dot{H}_n(\epsilon) = -\tthalf k H_n(0)$ with $\epsilon \rightarrow 0^+$. This condition follows in the usual way; see the Appendix. 

The original AdS/Yang-Mills (YM) model contained three parameters: $l$, a physically irrelevant parameter representing the AdS curvature; $\lambda$, representing the slope of the mass spectrum; and $\gamma$, representing the transition from the UV to the IR behavior.  None of the results depended upon $l$.  The lack of a UV cutoff in that model resulted in an instability manifested by an infinite action and a tachyon.  The AdS/YM model which includes the Randall-Sundrum 3-branes replaces the parameter $l$ with the parameter $l=1/k$ which is relevant: it shifts the variable $z$ by an amount $l$, and it enters into the Schr\"odinger equation for the glueball mass spectrum. It is related to the tension on the 3-branes. It also renders the action finite, hence we do not expect a tachyon.

Figure 6 shows a best fit to the glueball mass spectrum as calculated on the lattice.  The best-fit parameters are: $\sqrt{\lambda} = 581.8$ MeV, $k = 0.951 \sqrt{\lambda}$, and $\gamma = 0.2725$.  Not only is the linearity achieved at large $n$ but also the departure from linearity for the ground state $n=1$.  
\begin{figure}[h]
\center{\includegraphics[width=290pt]{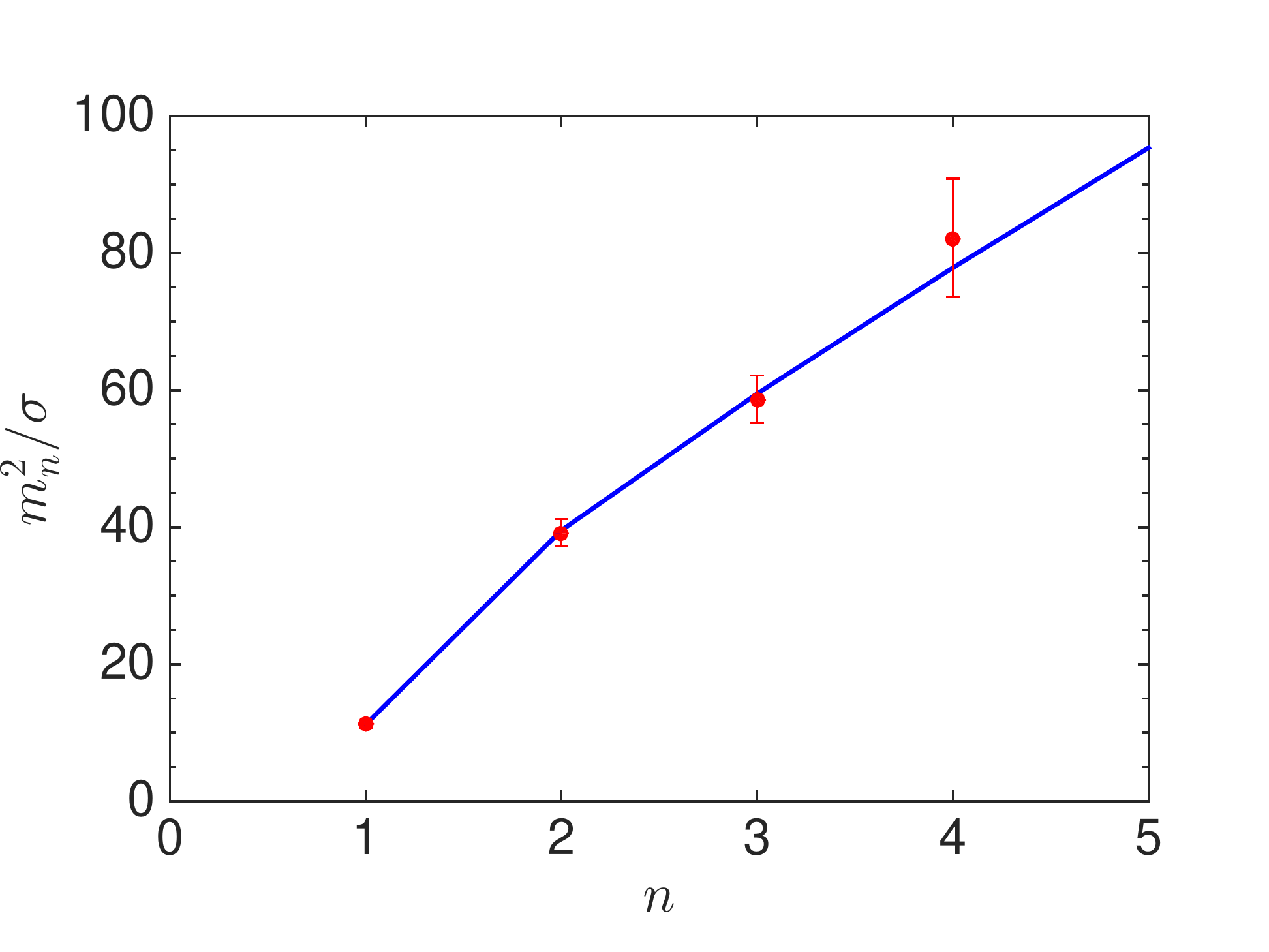}}
\caption{Glueball mass spectrum calculated with the AdS/YM model including the Randall-Sundrum 3-brane.  The best-fit parameters are given in the text.}
\label{spectra_log11}
\end{figure}
The potential entering the Schr\"odinger equation with these parameters is shown in Fig. 7.
\begin{figure}[h]
\center{\includegraphics[width=300pt]{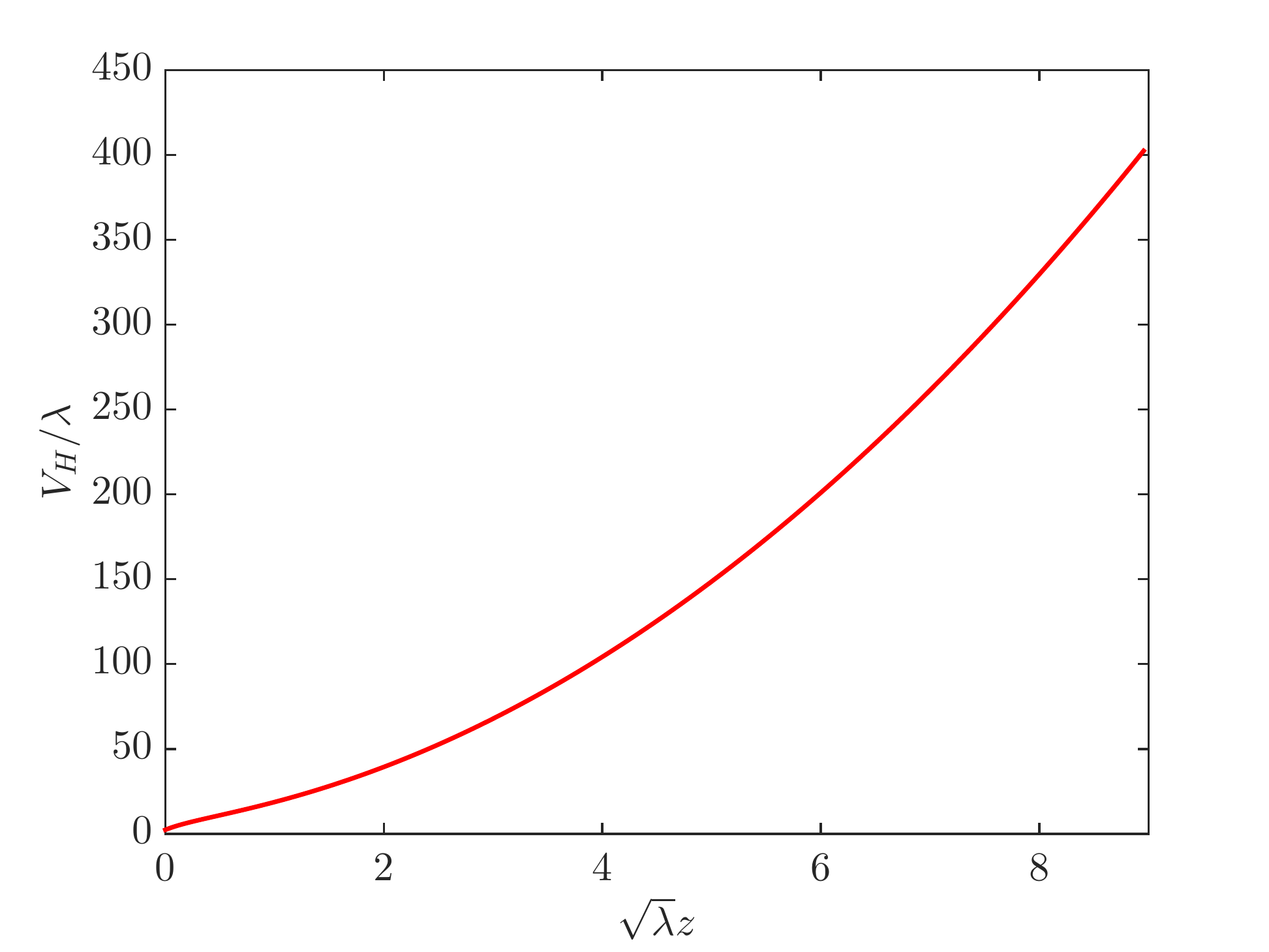}}
\caption{The Schr\"odinger potential for the AdS/YM model including the Randall-Sundrum 3-brane.}
\label{spectra_log22}
\end{figure}
Note that it does not have the $1/x^2$ behavior exhibited in Eqs. (\ref{pot1}) and (\ref{pot2}) as $x \rightarrow 0$ because of the UV cutoff $l=1/k$.  

%%%%%%%%%%%%%%%%%%%%%
\newpage
\section{Conclusion}
\label{conclude}

Our goal in this paper was to construct a fully dynamical AdS-type model that could reproduce the scalar glueball mass spectrum as calculated by lattice gauge theory.  The model should respect the gauge/gravity duality correspondence.  The first hurdle in doing so is the fact that there is always one more equation of motion for the background fields than there are independent fields.  This hurdle was overcome by using the superpotential method, whereby one function $U(G)$ determines the full potential $V(\phi,G)$ in the Lagrangian.  Then it is straightforward to parametrize $U(G)$ such that the correspondence is satisfied and that the scalar glueball spectrum has a linear trajectory for the radial excitations.  The theory has two parameters: $\lambda$, which arises as a constant of integration in the differential equations determining the background fields and which introduces an overall energy scale (there is no such scale in the Lagrangian), and a dimensionless parameter $\gamma$ which determines the scale at which the theory switches from the IR to the UV behavior.   Unfortunately any such theory has a tachyon, which indicates an instability.

The model is stabilized by merging the AdS approach with the Randall-Sundrum model.  In that model a cosmological constant is introduced in the bulk and 3-branes are introduced at $y=0$ and $y=L$.  The length $L$ can be taken as large as desired and hence is an irrelevant parameter.  It introduces one new parameter, $k$, with dimensions of energy which provides a UV cutoff and renders the action finite.  This eliminates the tachyon.  An excellent fit to the 4 scalar glueball states calculated by lattice gauge theory is obtained.

The next step in our program is to study the theory at finite temperature.  Nonzero temperature will modify the background fields $\phi(z)$ and $G(z)$.  The outcome would be the equation of state and the transport coefficients, such as the shear and bulk viscosities and various relaxation times that appear in higher order viscous fluid dynamics.  One can also study the spectra of the $J^{PC} = 0^{-+}, 1^{+-}$ and $1^{--}$ states.  Finally, one may consider applying this approach to the full bottom-up AdS/QCD phenomenology.  There is much work yet to be done!  

\section*{Acknowledgments}
We thank T. Gherghetta for very insightful discussions. This work was supported by the U.S. DOE Grant No. DE-FG02-87ER40328.  A. D. is supported
by the National Science Foundation Graduate Research Fellowship Program under Grant No. 00039202.

\section*{Appendix A}

The Riemann-Christoffel tensor $R_{\alpha \beta \mu \nu}$ for the background metric has the following independent components.  All the rest are either related to these by the algebraic properties of $R_{\alpha \beta \mu \nu}$ or are zero.  These results apply when the metric is diagonal and depends only on the fifth-dimensional coordinate, here labeled $r$.  These results are taken from \cite{Springer2008}.  The formulas from Ref. \cite{Springer2008} use Weinberg's \cite{Weinberg} definition of the Riemann-Christoffel tensor, whereas the formulas in the text use Carroll's \cite{Carroll} definition.  They differ by an overall sign.  The formulas below use Carroll's definition:
\ba
R_{r0r0} &=& - \frac{1}{4} \frac{dg_{00}}{dr} \Dr \ln \left[ g^{rr} g^{00}
\left( \frac{dg_{00}}{dr} \right)^2 \right] \nonumber \\
R_{rxrx} = R_{ryry} = R_{rzrz} &=& - \frac{1}{4} \frac{dg_{xx}}{dr} 
\Dr \ln \left[ g^{rr} g^{xx}
\left( \frac{dg_{xx}}{dr} \right)^2 \right] \nonumber \\
R_{0x0x} = R_{0y0y} = R_{0z0z} &=& - \frac{1}{4} g^{rr} \frac{dg_{00}}{dr}
\frac{dg_{xx}}{dr} \nonumber \\
R_{xyxy} = R_{xzxz} = R_{yzyz} &=& - \frac{1}{4} g^{rr} 
\left( \frac{dg_{xx}}{dr} \right)^2
\ea
The diagonal elements of the Ricci tensor $R_{\mu\nu} = g^{\alpha\beta}
R_{\alpha\mu\beta\nu}$ are nonzero while the off-diagonal ones are zero.
\ba
R_{00} &=& - \frac{1}{2} g^{rr} \frac{dg_{00}}{dr} \Dr \ln \left[ \root
g^{rr} g^{00} \frac{dg_{00}}{dr} \right] \equiv g_{00} F_0 \nonumber \\
R_{xx} = R_{yy} = R_{zz} &=& - \frac{1}{2} g^{rr} \frac{dg_{xx}}{dr} 
\Dr \ln \left[ \root g^{rr} g^{xx} \frac{dg_{xx}}{dr} \right] \equiv g_{xx} F_x \nonumber \\
R_{rr} &=& - \frac{3}{4} g^{xx} \frac{dg_{xx}}{dr} \Dr \ln \left[ g^{rr} g^{xx}
\left( \frac{dg_{xx}}{dr} \right)^2 \right] \nonumber \\
&-& \frac{1}{4} g^{00} \frac{dg_{00}}{dr} \Dr \ln \left[ g^{rr} g^{00}
\left( \frac{dg_{00}}{dr} \right)^2 \right] \equiv g_{rr} F_r 
\ea
The curvature $R = R^{\lambda}_{\;\;\lambda}$ is given as follows.
\be
R = F_0 + 3F_x + F_r
\ee

\section*{Appendix B}

Consider the differential equation
\be
- \ddot{H} + [v(z) - \lambda \delta(z)] H = m^2 H
\ee
where $v(z)$ is a smooth, even function of $z$ and $-\infty < z < \infty$.  Integrate this equation once from $-\epsilon$ to $\epsilon$.  In the limit $\epsilon \rightarrow 0$ it gives
 $-[\dot{H}(\epsilon) - \dot{H}(-\epsilon)] = \lambda H(0)$.  Since $H$ is continuous at 0 and is an even function it follows that $\dot{H}(-\epsilon) = - \dot{H}(\epsilon)$, hence $\dot{H}(\epsilon) = - \thalf \lambda H(0)$.

This can be thought of as a differential equation to be solved for $z \ge 0$ due to the symmetry.  It may be useful to represent the $\delta$-function as either a square well from $-a$ to $a$ or as a Gaussian of width $\sigma$ centered at 0.  Now it is clear that $\dot{H}(0) = 0$, so that integration from 0 to $\epsilon$ yields $\dot{H}(\epsilon) = - \thalf \lambda H(0)$, the same as before, as it must be.

\end{document}